\documentclass[12pt]{article}
\usepackage[normalem]{ulem}
\usepackage{geometry}
\usepackage{amsmath}
\usepackage{amssymb,epsfig,subfigure}
\usepackage[usenames, dvipsnames]{color}
\usepackage{graphicx}
\numberwithin{equation}{section}

\newtheorem{lemma}{Lemma}

\newcommand{\mult}{\operatorname{mult}}

\newcommand{\be}{\begin{equation}}
\newcommand{\ee}{\end{equation}}
\newcommand{\ba}{\begin{aligned}}
\newcommand{\ea}{\end{aligned}}

\def\m1{\left(-1\right)^{F_i}}

\makeatletter
\def\sla@#1#2#3#4#5{{%
  \setbox\z@\hbox{$\m@th#4#5$}%
  \setbox\tw@\hbox{$\m@th#4#1$}%
  \dimen4\wd\ifdim\wd\z@<\wd\tw@\tw@\else\z@\fi
  \dimen@\ht\tw@
  \advance\dimen@-\dp\tw@
  \advance\dimen@-\ht\z@
  \advance\dimen@\dp\z@
  \divide\dimen@\tw@
  \advance\dimen@-#3\ht\tw@
  \advance\dimen@-#3\dp\tw@
  \dimen@ii#2\wd\z@  \raise-\dimen@\hbox to\dimen4{%
    \hss\kern\dimen@ii\box\tw@\kern-\dimen@ii\hss}%
  \llap{\hbox to\dimen4{\hss\box\z@\hss}}}}
\def\slashed#1{%
  \expandafter\ifx\csname sla@\string#1\endcsname\relax
    {\mathpalette{\sla@/00}{#1}}%
  \else
    \csname sla@\string#1\endcsname
  \fi}
\makeatother

\begin{document}


\thispagestyle{empty}
\begin{flushright}\footnotesize
\texttt{UCSB Math 2011-09}\\
\texttt{IPMU11-0107}\\
\texttt{NSF-KITP-11-110}\\
\texttt{KCL-MTH-11-13}\\
\vspace{.5cm}
\end{flushright}

\renewcommand{\thefootnote}{\fnsymbol{footnote}}
\setcounter{footnote}{0}

\begin{center}
{\Large\textbf{\mathversion{bold} Tate's algorithm and F-theory}\par 
}

\vspace{1cm}

\textrm{Sheldon Katz$^1$, David R. Morrison$^{2,3,4}$,
 Sakura Sch\"afer-Nameki$^{5,6}$, and James Sully$^3$}

\vspace{.5cm}

\textit{$^1$ Department of Mathematics\\
University of Illinois at Urbana-Champaign\\
1409 West Green Street, Urbana, IL 61801, USA}

\vspace{.1cm}

\textit{$^2$ Department  of Mathematics\\
 University of California, Santa Barbara, CA 93106, USA}

\vspace{.1cm}

\textit{$^3$ Department of Physics\\
 University of California, Santa Barbara, CA 93106, USA}

\vspace{.1cm}

\textit{$^4$ Institute for the Physics and Mathematics of the Universe,
University of Tokyo\\ 5-1-5 Kashiwanoha,
Kashiwa, 277-8583, Japan}

\vspace{.1cm}

 \textit{$^5$ Kavli Institute for Theoretical Physics\\
 University of California, Santa Barbara, CA 93106, USA}

\vspace{.1cm}

\textit{$^6$ Department of Mathematics, King's College\\
University of London, The Strand, WC2R 2LS London, England}

\bigskip


\par\vspace{1cm}

\textbf{Abstract}\vspace{5mm}
\end{center}

\noindent
The ``Tate forms" for  elliptically fibered Calabi-Yau manifolds are reconsidered in order to determine  their general validity. 
We point out that there were some implicit assumptions made in the original derivation of these ``Tate forms'' from  the Tate algorithm.
By a careful analysis of the Tate algorithm itself, we deduce that the ``Tate forms" (without any futher divisiblity assumptions)
do not hold in some instances and have to be replaced by a new type of ansatz. Furthermore, we give examples 
in which the existence of a ``Tate form" can be globally obstructed, i.e., the change of coordinates does not extend globally to
sections of the entire base of the elliptic fibration. 
These results have implications both for model-building and for the exploration
of the landscape of F-theory vacua.

\vspace*{\fill}

\setcounter{page}{1}
\renewcommand{\thefootnote}{\arabic{footnote}}
\setcounter{footnote}{0}

 \newpage

\tableofcontents

\newpage

\renewcommand{\topfraction}{.9}
\renewcommand{\bottomfraction}{.8}

\section{Introduction}

F-theory 
offers an important complement to traditional, perturbative
compactifications of string theory.  F-theory utilizes the mathematical
theory of elliptic fibrations to produce compactifications of type IIB
string theory which fully exploit and manifest the $\operatorname{SL}(2,\mathbb{Z})$
S-duality of that theory.  In particular, F-theory compactifications {generically}
have no weak coupling regime  and must be
studied non-perturbatively.

Originally proposed as 8D effective theories
\cite{Vafa:1996xn}, 
F-theory models were soon extended to 6D \cite{FCY1,FCY2,geom-gauge}
and 4D \cite{Bershadsky:1996gx}, 
and extensively studied in the late 1990's. (See \cite{tasi,Denef:2008wq}
for reviews which include discussions of F-theory.)
The past several years have seen a significant revival of
the study of F-theory models in 4D, beginning with
\cite{Donagi:2008ca,Beasley:2008dc,Hayashi:2008ba}
and reviewed in {\cite{Weigand:2010wm}}. Moreover, F-theory has played an
important r\^{o}le in the study of the ``landscape'' of string
compactifications, both in four dimensions 
\cite{Grana:2005jc,Douglas:2006es,Denef:2007pq}
and in recent work in six dimensions 
\cite{Kumar:2009ae,mapping,global6D,Taylor:2011wt}.

Since its very beginning, F-theory has been closely tied to the
mathematics of elliptic fibrations, a subject which got its
start with work of Kodaira \cite{MR0184257} 
nearly
fifty years ago.
Kodaira's original classification of one-parameter families
of elliptic curves was extended to a number-theoretic setting
by N\'eron \cite{MR0179172}, and  was analyzed in an algorithmic
way by Tate \cite{MR0393039}, whose approach can be applied in
both the number theory and algebraic geometry contexts.

The Kodaira classification and the Tate algorithm have become important
tools in analyzing F-theory models.  One of the early papers
to use the Tate algorithm for this purpose \cite{geom-gauge} derived
a collection of {{ans\"atze}} which can be used for constructing F-theory
models with a specified gauge group.  (It has become common to refer
to models in one of the forms
 given in \cite{geom-gauge} as being in ``Tate
form''.) 

 Recent work to construct supersymmetric grand unified theories (GUTs) in F-theory 
have relied heavily on the Tate form of the singularity, in particular in the context of Higgs bundles and spectral cover 
constructions for these models \cite{Donagi:2009ra, Marsano:2009gv}. In this context, we consider an elliptically fibered Calabi-Yau fourfold, 
with singularity over a divisor in the base. The singularity type determines the gauge
group realized on this codimension one locus and the ``Tate form" is a very useful starting point for analysing the dynamics of this gauge theory. 
For instance, in the case of $SU(5)$, the ``Tate form" is precisely an unfolded $E_8$ singularity, which from the point of view of the 
gauge theory has the interpretation that the $SU(5)$ results from Higgsing an $E_8$ gauge theory. 
It is therefore interesting, also from a physics point of view, to
know whether the ``Tate form" can be achieved in general (which in the case of $SU(5)$ is indeed possible) and whether it holds up
globally (we give some discussion in appendix C).

Unfortunately, in the derivation of the ``Tate form" ans\"atze  in \cite{geom-gauge}  the assumptions
made were not spelled out very clearly\footnote{Potential problems were pointed out, for example,
in footnote 2 of \cite{braneCY}.}.  We re-analyze the validity of
these ``Tate forms'' in this paper. 

Our main conclusion is that, with the exception of one new ansatz which
must be introduced for certain $Sp$ groups, 
the ``Tate forms'' of \cite{geom-gauge} hold up fairly
well, as long as {{
\begin{itemize}
\item[ (1)] we avoid certain matter representations 
(such as the $2$-symmetric representation of $SU(m)$) which
are associated with singularities in components of the discriminant 
locus of the elliptic fibration \cite{Sadov:1996zm}
\item[(2)] we avoid gauge groups $SU(m)$ with 
$6\le m\le 9$, $Sp(n)$ with 
$n=3, 4$,
and $SO(\ell)$ with 
$\ell = 13, 14$.
\end{itemize}
}}

These restrictions will not seem too confining for those constructing
explicit GUT models using F-theory, since  no change is needed for
the widely-studied GUT groups $SU(5)$, $SO(10)$, or $E_6$.  (The 
``dangerous'' matter representations {{are}} also avoided in standard GUT scenarios.)
For those wishing to explore the entire landscape of F-theory vacua,
however, our results are a signal that more work must be done before
truly systematic studies can be made.  In fact, some of that work 
has been completed and is being reported on in a companion paper to
this one \cite{matter1}.

The Tate algorithm proceeds by a sequence of coordinate changes adapted to the
geometry of the fibration.  We wish to
emphasize that the coordinate changes necessary to write the fibration in 
Tate form may only be defined locally on the base.
The main result of the paper is that, by a careful analysis of the algorithm, we find that there are
local coordinate 
changes  which bring the fibration into either ``Tate form''
or a new ansatz which we introduce in this paper (except for classical
groups of certain rank, or cases which involve certain exotic matter
representations).

Another result of this paper is the construction of
a simple explicit example for which we  demonstrate that
the required coordinate change cannot be defined globally.
This aspect of Tate's algorithm has not been taken into account in the recent
constructions of GUT models via F-theory, and it seems likely that the added
flexibility provided by treating Tate's algorithm only locally will
allow the construction of new models.  We leave such constructions
 for future work.

The outline of our paper is as follows.
In section 2, we  fix our notation for the Weierstrass form and
introduce the first coordinate change used in Tate's algorithm. 
In section 3 we show that the coordinate changes leading to Tate form may 
fail to be defined globally, requiring different coordinate changes in
different local coordinate charts. 
In sections 4-7 we discuss in turn all singularity types and spell out in 
detail the steps that are required to achieve Tate form. That is, we
go through Tate's algorithm carefully, paying close attention to issues
that arise in codimension two and greater.
We determine normal forms for local equations for the classical
groups $SU(m)$, $Sp(n)$, and $SO(\ell)$ of sufficiently large rank
by using an induction argument for the  $I_m$ and $I_m^*$ cases.
Sections 4.10 and 6.5 give concise summaries of our results for these cases.
In section 8, we state our conclusions.

For completeness, we reproduce the main table with the ``Tate form''
ans\"atze from \cite{geom-gauge} in Appendix~\ref{app:tateform}.
We also have Appendix~\ref{app:ufd} on unique factorization, and {{Appendix~\ref{app:C} where we spell out 
some sufficient 
conditions for global obstructions for $SU(5)$.}}


\section{Normal forms for Weierstrass equations}\label{sec:normal}

We consider an elliptically fibered Calabi--Yau manifold $Y$ (of arbitrary
dimension) with Weierstrass equation
\be\label{eq:Weier}
y^2 = x^2 + fx + g \,,
\ee
and let $B$ denote the  base of the fibration.  Each local 
factor\footnote{Two compact Lie groups are said to be {\em locally
isomorphic}\/ when they have isomorphic Lie algebras.  The gauge
group of these theories is built from various ``local'' factors
(determined by the corresponding Lie algebra) by forming the product
group from the factors and then taking a quotient by a finite group,
if necessary.} $G$ of
the nonabelian part of
the gauge group of the corresponding F-theory model is associated with
a divisor $S=S_G \subset B$, over which
a singularity is located which enhances the gauge group.  We assume
that 
 each such divisor $S$ is nonsingular, which implies that on any sufficiently
small (Zariski)
open set of $S$, the coordinate ring is a unique factorization domain
\cite{MR0103906}.
We will use the unique factorization property repeatedly in our analysis.

Our focus is on one chosen divisor $S$, but our analysis can be applied
to any gauge-symmetry-enhancing divisor on $B$ (as long as it is nonsingular).
Note, however, that at an intersection point between two or more such
divisors, the coordinate changes dictated
by Tate's algorithm for each of the divisors may be different.

If we restrict to a sufficiently small Zariski open set  $U\subset B$,
that is, a sufficiently small set 
$U$ whose complement $W=B-U$ is defined by polynomial equations,
the restriction
$S|_U$ will have a local defining equation of the form $\{z=0\}$, and
we can expand the Weierstrass coefficients $f$ and $g$ as power series
in $z$
\be
f=  \sum_{i}f_i z^i \,,\qquad 
g=  \sum_i g_i z^i \,.
\ee
The coefficients in this expansion are algebraic functions on $U$ but
they may have poles on $W=B-U$; different expressions may be needed for
different open sets.  Moreover, the leading non-zero coefficients, when
restricted to $S$, are well-defined, but the higher terms in the sequence
may not be well-defined.  We will comment on these issues further when they
arise in our computation.

\begin{table}
{\footnotesize
\begin{center}
\begin{tabular}{|c|c|c|c|c|c|c|} \hline
&$\operatorname{ord}_{S}(f)$&$\operatorname{ord}_{S}(g)$
&$\operatorname{ord}_{S}(\Delta)$
&singularity&local gauge group factor\\ \hline\hline
$I_0$&$\ge0$&$\ge0$&$0$&none
& --\\ \hline
$I_1$&$0$&$0$&$1$&none& -- \\ \hline
$I_2$&$0$&$0$&$2$&$A_1$
&$SU(2)$\\ \hline
$I_m$, $m\ge1$&$0$&$0$&$m$&$A_{m-1}$
&$Sp([\frac m2])$ or $SU(m)$\\ \hline
$II$&$\ge1 $&$   1  $&$    2 $&  none
&--\\ \hline
$III$&$  1 $&$   \ge2 $&$   3 $&$  A_1$
&$SU(2)$\\ \hline
$IV$&$ \ge2 $&$  2  $&$    4 $&$  A_2$
&$Sp(1)$ or $SU(3)$\\ \hline
$I_0^*$&$\ge2$&$\ge3$&$6$&$D_{4}$
&$G_2$ or $SO(7)$ or $SO(8)$\\ \hline
$I_m^*$, $m\ge1$&$2$&$3$&$m+6$&$D_{m+4}$
&$SO(2m+7)$ or $SO(2m+8)$\\ \hline
$IV^*$&$\ge3$&$  4$  &$  8$&$   E_6$
&$F_4$ or $E_6$\\ \hline
$III^*$&$  3 $&$   \ge5 $&$   9 $&$  E_7$
&$E_7$\\ \hline
$II^*$&$ \ge4$&$   5   $&$   10 $&$  E_8$
&$E_8$\\ \hline
non-minimal&$\ge4$&$\ge6$&$\ge12$&non-canonical&--\\ \hline
\end{tabular}
\end{center}
\medskip
\caption{Kodaira's classification of singular fibers and gauge groups}\label{tab:kodaira}
}
\end{table}

The discriminant of the elliptic fibration is
\be
\Delta = 4 f^3 + 27 g^2 \,,
\ee
and Kodaira's analysis \cite{MR0184257} (the results of which are
reproduced in Table~\ref{tab:kodaira}) determines 
the general singularity type along $S$
in terms of
the orders of vanishing of $f$, $g$, and $\Delta$.  To determine the 
local contribution $G$ to the gauge group,
one must also use the part of Tate's analysis \cite{MR0393039}
which specifies the
monodromy of the exceptional curves along $S$.  The various possibilities
for each Kodaira type are exhibited in the final column of Table~\ref{tab:kodaira}.  We will discuss the conditions on the equation which determine the
monodromy (and local gauge group) when we come to them in the algorithm.

Expanding $\Delta$ in $z$, the leading terms are
\be\label{Delta2}
\Delta = \left(4 f_0^3+27 g_0^2\right)
+ \left(12 f_1 f_0^2+54 g_0 g_1\right)z+O\left(z^2\right) \,.
\ee
As described in  Appendix~\ref{app:ufd}, because $S$ is nonsingular,
we can find a function $u_0$ (possibly after
shrinking $U$) such that
\be
f_0 =-\frac13 u_0^2 +O(z)\,, \qquad
g_0 =\frac2{27} u_0^3 +O(z)\,.
\ee

We replace $f_0$ by $-\frac13u_0^2$ and $g_0$ by $\frac2{27}u_0^3$, 
modifying the higher coefficients as necessary.
Now following Tate, we make a change of coordinates
\be
(x,y) \mapsto (x+\frac13u_0,y)
\ee
which transforms the defining equation to
\be 
y^2 = x^3+u_0x^2 
 +(f_1z+f_2z^2+\cdots)x+
(g_1+\frac13u_0f_1)z+
(g_2+\frac13u_0f_2)z^2+
(g_3+\frac13u_0f_3)z^3+\cdots
\ee
As can be seen in Appendix~\ref{app:tateform},
this is ``Tate form'' for type $I_1$.
To simplify later formulas, we set $\widetilde{g}_j=g_j+\frac13u_0f_j$,
and write the equation in the form
\be \label{eq:tateI1}
y^2 = x^3+u_0x^2 
 +(f_1z+f_2z^2+\cdots)x+
(\widetilde{g}_1z+
\widetilde{g}_2z^2+
\widetilde{g}_3z^3+\cdots)
\ee

More precisely, here and in the rest of this paper, by ``Tate
form'' we mean the more compact version given in 
Table~\ref{tab:tatealgcompact}, where an entry ``$\infty$'' means that
a coefficient has been set to zero.


\section{Global obstructions to Tate form: An example}\label{sec:obstruction}


In this section we give an example which shows that the key coordinate
change used in the preceding section to bring a Weierstrass equation into a Tate-type normal form may only be possible locally.

For simplicity we consider a two-dimensional base $\mathbb{P}^1\times \mathbb{P}^1$.  
Line bundles on this variety are denoted by 
$\mathcal{O}_{\mathbb{P}^1\times\mathbb{P}^1}(a,b)$, labeled by their
bi-degree $(a,b)$.
As we will see,
the important property for the
example is that 
$H^1(\mathbb{P}^1\times \mathbb{P}^1,\mathcal{O}_{\mathbb{P}^1\times\mathbb{P}^1} (0,-2))\ne\{0\}$.
We use $[x_0,x_1]$ and $[y_0,y_1]$ as homogeneous coordinates on the
two $\mathbb{P}^1$ factors.

Consider a fibration determined by 
Weierstrass coefficients
$f = \frac13a(x)b(x)y_0^4y_1^4$
and $g = -\frac2{27}a(x)b(x)^2y_0^3y_1^9$, where $a(x_0,x_1)$ and
$b(x_0,x_1)$ are homogeneous polynomials of degree $4$. 
The discriminant is given by
\[\Delta = 4f^3+27g^2 = \frac4{27}a(x)^2b(x)^3y_0^6y_1^{12}\left(a(x)y_0^6+b(x)y_1^6\right)\]
Let $S=\{z=0\}$ be the divisor with equation 
 $z = a(x)y_0^6+b(x)y_1^6 $.

By the standard Tate procedure outlined in the preceding section we can find functions
\[u_0= -\frac92\frac gf =\frac{b(x)y_1^5}{y_0}\]
and
\[u_1 = \frac23\frac{f^2}g=\frac{-a(x)y_0^5}{y_1} \, .\]
defined respectively on the open sets $ U_0 = \lbrace y_0 \neq 0\rbrace$ and $ U_1 = \lbrace y_1 \neq 0\rbrace$ 
such that on 
each $U_i\cap S$ we have $f = -(1/3) u_i^2+O(z)$ and $g=(2/27) 
u_i^3 + O(z)$.  Note that on
$(U_0 \cap U_1)|_S$ we have that
\begin{equation} \label{eq:difference}
 u_0|_{U_0\cap U_1} - u_1|_{U_0\cap U_1} =
\left. \frac{b(x)y_1^6 + a(x)y_0^6}{y_0y_1}\right|_{U_0\cap U_1},
\end{equation}
which implies that $u_0|_{U_0\cap U_1\cap S} = u_1|_{U_0\cap U_1\cap S}$.
We define $\tilde u := u_i |_S $.

We will show there does not exist any global section $u$ such that 
$u|_{U_0}\equiv u_0$ modulo $z$ and $u|_{U_1}\equiv u_1$ modulo $z$. 
Our argument relies on the exact sequence of sheaves
\begin{align}
0 \rightarrow \mathcal{O}_{\mathbb{P}^1 \times\mathbb{P}^1}\left(0,-2\right) 
\stackrel{\cdot z}{\longrightarrow} \mathcal{O}_{\mathbb{P}^1 \times\mathbb{P}^1}\left(4,4\right)  
\longrightarrow \mathcal{O}_{S}\left(4,4\right) \rightarrow 0
\, ,
\label{eq:exseq}
\end{align}
where the first map is multiplication by the equation $z$ of $S$.
Let $\mathcal{U} = \{U_0,U_1\}$ be the open cover of the base defined
by our open sets.
We have $\tilde u\in H^0(\mathcal{U},\mathcal{O}_{S}\left(4,4\right)),$
where here and in the sequel the notation $H^i(\mathcal{U},\ldots)$ 
emphasizes that we are computing \v{C}ech cohomology for the cover $\mathcal{U}$.

Let us
compute the image of $\tilde u$ under the coboundary map
\[ H^0(\mathcal{U},\mathcal{O}_{S}\left(4,4\right)) \to
H^1(\mathcal{U},\mathcal{O}_{\mathbb{P}^1 \times\mathbb{P}^1}\left(0,-2\right)).\]
To compute the image, we use the function $u_j$ as a lift
of $\tilde u|_{U_j\cap S}$ to $U_j$.  The difference $u_0-u_1$ then
maps to zero when restricted to $S$, and so is in the image of multiplication
by $z$.  We compute the pre-image using \eqref{eq:difference}:
\[\left. \frac{u_0-u_1}{z}\right|_{U_0\cap U_1}
= \left.\frac1{y_0y_1}\right|_{U_0\cap U_1}\]
so under the coboundary map, $\tilde u$ maps to
\[ \left(\frac1{y_0y_1}\right)\in H^1(\mathcal{U},\mathcal{O}_{\mathbb{P}^1 \times\mathbb{P}^1}\left(0,-2\right)).\]
But it is easy to see that this is a non-trivial class in that cohomology
group: in fact, it generates it.

Thus, since $\tilde u$ has a nonzero image in 
$H^1(\mathcal{O}_{\mathbb{P}^1 \times\mathbb{P}^1}\left(0,-2\right))$,
it cannot lie in the image of the map
\[ H^0(\mathcal{O}_{\mathbb{P}^1 \times\mathbb{P}^1}\left(4,4\right))
\to H^0(\mathcal{O}_{S}\left(4,4\right)).\]
That is, there is no global section $u$ of $\mathcal{O}_{\mathbb{P}^1 \times\mathbb{P}^1}\left(4,4\right)$ which restricts to $u_0$ modulo $z$ on $U_0$
and restricts to $u_1$ modulo $z$ on $U_1$.  Thus, our coordinate
change cannot be made globally.

We have chosen a particularly simple $f$ and $g$ for
expository purposes; the construction is much more general.
In fact, our example is too simplistic: 
the resulting Calabi-Yau has physically unacceptable singularities over the 
lines $y_0=0$ and $y_1=0$ in the base.  
However, we can 
alter the equations of $f$ and $g$ by adding generic multiples of $z$.  This
will remove the singularities but will not alter the nonvanishing element of
$H^1(\mathcal{U},\mathcal{O}_{\mathbb{P}^1 \times\mathbb{P}^1}\left(0,-2\right))$ computed above.

Note that this example works equally well at higher order in the vanishing of the discriminant. For example, in the case $I_2$, we can use
\begin{align}
f &\equiv \frac13a(x)b(x)y_0^4y_1^4 + y_1 f_1 z \nonumber\\
g &\equiv -\frac2{27}a(x)b(x)^2y_0^3y_1^9 + y_0^3\left(\frac{1}{3} a(x) f_1 y_0^2 +\frac{1}{27} a(x)b(x) y_1^3\right)z 
\, ,
\end{align}
as Weierstrass coefficients,
in which case we face the same non-existence of the required global section.

In appendix C we repeat this analysis for the case of $SU(5)$ singularities, which are particularly interesting for 
$SU(5)$ GUT model-building. Rather than present an explicit example, we derive one set of sufficient conditions to construct a globally obstructed $SU(5)$ fibration. There are surely many other ways to construct analogous examples.


\section{The $I_m$ case}


We return to a general base manifold, and go through Tate's algorithm
step by step.  At each step, Tate's algorithm specifies a coordinate
change to be made, and in the original formulation these coordinate
changes can involve rational functions on the base (with denominators 
allowed).  Our goal here is to find a version of the algorithm in
which the coordinate changes can be made without using functions with
poles on the base.

We  first treat the case of $I_m$.
According to Kodaira, to be in the $I_m$ branch of the classification
we must have
 $z\not|u_0$.  As noted in section~\ref{sec:normal},
 when $m=1$ the normal form
\eqref{eq:tateI1} coincides with ``Tate form'' for $I_1$.



\subsection{Step 1}

For an equation of the form \eqref{eq:tateI1}, the leading terms of the
discriminant can be written
\be
\Delta = 4u_0^3\widetilde{g}_1z+O\left(z^2\right) \,.
\ee
Assuming that $m\ge2$ so that $z^2|\Delta$, we have
\be
\widetilde{g}_1 =0 + O(z)\,.
\ee
Thus, we may absorb $\widetilde{g}_1$ into $\widetilde{g}_2$ by 
adjusting the coefficients, and can assume that $\widetilde{g}_1$
has been set to $0$.  In this case,  the defining equation becomes
\be
y^2 = x^3+u_0x^2 
 +(f_1z+f_2z^2+\cdots)x+
(  \widetilde{g}_2z^2+
\widetilde{g}_3z^3+\cdots) \,.
\ee
This is ``Tate form'' for type $I_2$.


\subsection{Step 2}

At the next order the discriminant is
\be
\Delta = u_0^2(4u_0\widetilde{g}_2-f_1^2)z^2+O\left(z^3\right) \,.
\ee
As explained in Appendix~\ref{app:ufd}, if the leading term vanishes 
(i.e., $m\ge3$) then we can
find functions $s_0$ and $\mu$ such that $\mu|_S$ is square-free, and a 
function $t_1$, such that
\be
u_0=  \frac14\mu s_0^2 +O(z) \,,\qquad
f_1 = \frac12\mu s_0  t _1 +O(z) \,.
\ee

The monodromy condition for $I_m$ is tested by asking whether
$u_0|_S$ has a square root or not: this determines whether the
local gauge group factor is $SU(m)$ or $Sp([\frac m2])$.  
With our notation, this amounts to asking
whether $\mu|_S$ has any zeros or not.  We will assume that $\mu$ has been
chosen so that $\mu\equiv1$ if $\mu|_S$ has no zeros, and this will be our
criterion for monodromy.  That is, in the $I_m$ case the local factor $G$
of the gauge group
will be $SU(m)$ if $\mu\equiv1$, and will be $Sp([\frac m2])$
if $\mu|_S$ has zeros.

Now we can solve for $\widetilde{g}_2$ as
\be
\widetilde{g}_2 =\frac{1}{4} \mu t _1^2 +O(z)\,.
\ee
We replace $u_0$, $f_1$, and $\widetilde{g}_2$ by $\frac14\mu s_0^2$,
$\frac12\mu s_0  t _1$, and $\frac{1}{4} \mu t _1^2$ respectively,
and adjust the other coefficients accordingly.

In order to put this into Tate form, we would like to make the
substitution $x\mapsto x - t_1z/s_0$.  However, 
we cannot do this near zeros of $s_0$.  

Thus, the ``Tate form'' for $I_3$ which was described in \cite{geom-gauge}
cannot be
 achieved.  We introduce a new ansatz for $I_3$ (and will eventually
extend this to all $I_{2n+1}$ cases):
\be \label{eq:I3}
y^2 = x^3+\frac14\mu s_0^2 x^2 
 +(\frac12\mu s_0t_1z+f_2z^2+f_3z^3+\cdots)x+
(\frac1{4}\mu t_1^2 z^2+
\widetilde{g}_3z^3+\widetilde{g}_4z^4+\cdots)
\ee
Unlike the ``Tate forms'' presented in \cite{geom-gauge}, this ansatz cannot
be described purely in terms of the vanishing of certain coefficients
in an expansion, but involves a particular relationship among leading
terms in the expansion of the coefficients of $x^2$, $x^1$, and $x^0$ in
the equation.

\subsection{Step 2 without monodromy}

In the case of $I_3$ with no monodromy (i.e., the case of $SU(3)$),
we have $\mu\equiv1$, and there {\em is}\/ a change of
coordinates which puts this into Tate form:
\be
(x,y)\mapsto (x,y+\frac12s_0x+\frac12 t_1z) \,,
\ee
which yields the equation
\be
 y^2 + s_0xy+t_1zy 
= x^3
 +(f_2z^2+f_3z^3+\cdots)x+
( \widetilde{g}_3z^3+\widetilde{g}_4z^4+\cdots) \,.
\ee
This is ``Tate form'' for $SU(3)$ (i.e., $I_3$ with no monodromy).

\subsection{Step 3}\label{subsec:step3}

The discriminant at the next order is
\be
\Delta =
\frac{1}{16} \mu^3s_0^3   \left(s_0^3\widetilde{g}_3 - s_0^2t_1f_2 - t
   _1^3\right)z^3+O\left(z^4\right) \,.
\ee
We now assume in addition that $z^4|\Delta$, which can be achieved by
\be
t_1 = -\frac13s_0 u_{1}+O(z) \,,\qquad 
\widetilde{g}_3 =-\frac13  u _{1}f_2-\frac1{27}u _{1}^3+O(z) \, ,
\ee
using Lemma~\ref{lem:4} {{in appendix B}}.
We replace $t_1$ and $\widetilde{g}_3$ by $-\frac13s_0 u_{1}$
and $-\frac13  u _{1}f_2-\frac1{27}u _{1}^3$ respectively 
in (\ref{eq:I3}),
and adjust the other coefficients accordingly.
Our equation becomes
\be 
\ba
y^2 =& x^3+\frac14\mu s_0^2 x^2 
 +(-\frac16\mu s_0^2u_{1}z+f_2z^2+f_3z^3+\cdots)x \cr
 &+
\frac1{36}\mu s_0^2u_{1}^2 z^2+
(-\frac13u_1f_2-\frac1{27}u_1^3)
z^3+\widetilde{g}_4z^4+\cdots)\,.
\ea\ee
This can be simplified with the change of coordinates
$(x,y)\mapsto (x+\frac13u_{1}z,y)$ which yields:
\be
y^2 = 
 x^3
+(u_0+u_1z + \cdots)x^2
 +((f_2+\frac1{3}u_{1}^2)z^2+f_3z^3+f_4z^4+\dots)x+
(\widehat{g}_4z^4+\widehat{g}_5z^5+\cdots)
\ee
where $u_0=\frac14\mu s_0^2$ as above, and
$\widehat{g}_j = \widetilde{g}_j +\frac13u_{1}f_{j-1}$.
Let $\widehat{f}_2=f_2+\frac13u_1^2$.
This is ``Tate form'' for $I_4$.  

In the case without monodromy we again have $\mu=1$ and there is a coordinate change 
\be\label{eq:NoMonoShift}
(x, y) \mapsto \left(x, y+ \frac12 s_0 x\right) \,
\ee
which puts the equation into the ``Tate form" for $SU(4)$ 
\be
y^2 +   s_0 x y = 
 x^3 
+(u_1z + \cdots)x^2
 +\left({{\widehat{f}_2}} z^2+f_3z^3+f_4z^4+\dots\right)x+
(\widehat{g}_4z^4+\widehat{g}_5z^5+\cdots) \,.
\ee
In summary for $I_4$, the ``Tate form" can be achieved with and without monodromy, yielding either $Sp(2)$ or $SU(4)$ gauge groups, respectively.


\subsection{Step 4}

Finally, the discriminant at order $z^4$  is
\be
\Delta = \frac{1}{16} \mu^2s_0^4   \left( \mu s_0^2 
   \widehat{g}_4- \widehat{f}_2^2\right)z^4+O\left(z^5\right) \,.
\ee
If $z^5|\Delta$,
as in step 2, this can be solved as follows.
Since $\mu|_S$ is square-free, we must have $(\mu s_0)|_S$ dividing 
$ \widehat{f}_2^2|_S$.  That is, there exists a function $t_2$ (possibly
after shrinking $U$) such that
\be
\widehat{f}_2 = \frac{1}{2} \mu s_0   t _2 +O(z)\,.
\ee
Then to satisfy the discriminant condition, we simply need
\be
\widehat{g}_4 = \frac14 \mu t _2^2 +O(z)\,.
\ee
We replace $\widehat{f}_2$ and $\widehat{g}_4$ by
$\frac{1}{2} \mu s_0   t _2$ and $\frac14 \mu t _2^2$ respectively,
and adjust the other coefficients accordingly.
The equation becomes
\be
y^2 = 
 x^3
+(\frac14\mu s_0^2+u_1z)x^2
 +(\frac12\mu s_0t_2z^2+f_3z^3+f_4z^4+\dots)x+
(\frac14\mu t_2^2z^4+\widehat{g}_5z^5+\cdots)
\ee
This is our new ansatz for $I_5$, and it again involves a particular relationship among leading
terms in the expansion of the coefficients of $x^2$, $x^1$, and $x^0$ in
the equation.
(The ``Tate form'' for $I_5$ without monodromy would require
the substitution $x\mapsto x-t_2z^2/s_0$,                                      
which is not  possible near zeros of $s_0$.)


\subsection{Step 4 without monodromy}

However, for $I_5$ with no monodromy (i.e., $SU(5)$), we can again
achieve Tate form by a different change of variables.  Since there is
no monodromy, $\mu\equiv1$ and
we can make the change of variables
\be
(x,y) \mapsto
(x, y+\frac12 s_0x + \frac12t_2z^2) \,,
\ee
which yields the equation
\be
y^2 + s_0 xy + t_2z^2y
=
 x^3
+u_1zx^2
 +(f_3z^3+f_4z^4+\dots)x+
(\widehat{g}_5z^5+\cdots)
\ee
This is ``Tate form'' for $SU(5)$.


\subsection{Induction}

In subsection~\ref{subsec:step3}, 
we showed how to obtain the Tate form up to $I_4$ (making no further assumptions about monodromy).
To extend this to higher $I_m$ singularities we will now make an inductive {{argument}}, starting with a Tate form for an $I_{2n}$ singularity with $n\ge2$, with monodromy. It will be useful to write this in terms of
\be 
y^2=x^3+ux^2+vx+w \,,
\ee
or expanded in terms of $z$
\begin{equation} \label{eq:TateI2n}
\ba
y^2= &x^3 + (u_0+u_1z+u_2z^2+ \dots+u_{n-1}z^{n-1}) x^2 
+ (v_n z^n + v_{n+1}z^{n+1} + \dots)x  \cr
&+  (w_{2n} z^{2n} + w_{2n+1}z^{2n+1} + w_{2n+2}z^{2n+2} + \dots) \,,
\ea
\end{equation}
assuming that the expansion of $u$ contains no terms divisible by $z^n$,
and also assuming
\be
z^n\ |\ v \quad \hbox{and} \quad z^{2n}\ |\ w\,.
\ee
Since we are assuming
the Kodaira type is $I_m$ for some $m\ge2n$, we should have $z \not|\ u$.
As in the earlier analysis, we assume that $u_0$ takes the form
$\mu s_0^2$ with $\mu|_S$ square-free (and $\mu\equiv1$ when there
is no monodromy).

To relate this to the Weierstrass form used earlier (\ref{eq:Weier}) we complete the 
cube
to
\be
y^2 = \left(x+\frac13u\right)^3 + \left(-\frac13u^2+v\right)\left(x+\frac13u\right)
+\left(\frac2{27}u^3-\frac13uv+w\right) \,,
\ee
identifying\footnote{From these formulas we see that if $z$ had divided $u$,
we would not be in Kodaira type $I_m$.}
\be
 f=-\frac13u^2+v\,,\qquad 
 g=\frac2{27}u^3-\frac13uv+w \,.
\ee
 It follows that
 the discriminant is
\begin{align} 
\Delta &= 4\left(-\frac13u^2+v\right)^3 + 27 \left(\frac2{27}u^3-\frac13uv+w\right)^2\\
&= 4u^3w-u^2v^2-18uvw+4v^3+27w^2.  \label{eq:terms}
\end{align}
The known order of vanishing of each of these terms is $2n$, $2n$,
$3n$, $3n$, $4n$. Since we are assuming $n\ge2$
\be \Delta = u^2(4uw-v^2) + O(z^{2n+2}).\ee

The type of condition we are now going to use is a condition which holds
in codimension two on the base.  We already have a divisor $z=0$ which
is codimension one, and we have been considering quantities like $u_i$,
$v_i$ or $w_i$ which might have zeros at a subvariety of $z=0$, i.e.,
in codimension two.  For any subvariety $\Sigma\subset \{z=0\}$,
we can ask about the multiplicity along $\Sigma$ of $f$, $g$, and
$\Delta$ and apply Kodaira's classification to determine the generic
singularity type along $\Sigma$.  If those multiplicities satisfy
\[ \mult_\Sigma(f)\ge4, \quad \mult_\Sigma(g)\ge6, 
\quad \mult_\Sigma(\Delta)\ge12\]
(i.e., we are in the ``non-minimal'' part of Kodaira's classification
in codimension two), then we can blow up $\Sigma$ and still have
a Calabi--Yau total space of the fibration; this implies that the 
low-energy spectrum has peculiar things such as
light tensors and, massless strings.
  Thus, we will exclude such elliptic fibrations from consideration.

To use this condition here, we assume that  $m$, the actual
order of vanishing of the discriminant, is at least $10$.  (Our current
value of $n$ is related to this by $m\ge2n$.)  In that case, if the
multiplicity of $f$ exceeds $2$ and/or the multiplicity of $g$ exceeds
$3$, then 
we are not in 
either 
 of the $I_m$ 
or the $I_{m-6}^*$ cases,
so we
must be in one of the exceptional or non-minimal cases.
However, the multiplicity of $\Delta$ necessarily 
increases along some codimension two subvariety $\Sigma\subset
\Delta$, so
$\Delta$ has multiplicity 
strictly
greater than $10$ along $\Sigma$.  
Therefore
the model would
be non-minimal in codimension two,
by Kodaira's classification
in Table~\ref{tab:kodaira}.
Since we are assuming that this 
{\em doesn't}\/ happen, the multiplicity of $f$ is at most $2$ and
the multiplicity of $g$ is at most $3$.  This implies that the
multiplicity of $u$ is at most $1$, and that is the condition we
will actually use.

So, assuming our equation is in the form of eq.~\eqref{eq:TateI2n},
we expand $\Delta$ as follows:
\begin{align} \Delta 
& = u_0^2(4u_0w_{2n}-v_n^2) z^{2n} 
\\ & \quad + \left(u_0^2(4u_0w_{2n+1}+4u_1w_{2n}-2v_nv_{n+1})
+ 2u_0u_1(4u_0w_{2n}-v_n^2)\right)z^{2n+1} 
\\ & \quad + {{O(z^{2n+2})}}.
\end{align}

We first assume that $z^{2n+1}$ divides $\Delta$.  
Then $(4u_0w_{2n}-v_n^2)|_{z=0}$ must be identically zero.  Writing
$u_0=\frac14\mu s_0^2$ with $\mu|_S$ square-free, we see that $\mu s_0$ must
divide $v_n$ modulo $z$.  That is, there exists a function $t_n$ such
that $v_n = \frac12\mu s_0 t_n+O(z)$ and it then follows that
$w_{2n}=\frac14\mu t_n^2+O(z)$.
We replace $v_n$ and $w_{2n}$ by $\frac12\mu s_0 t_n$ and $\frac14\mu t_n^2$,
respectively, and adjust the other coefficients accordingly resulting in
\be\label{InMonoNew}
\ba
&y^2 =x^3  
+ x^2 \left({\frac14\mu s_0^2}  + u_1 z + u_2z^2 + \cdots  
+u_{n-1}z^{n-1}
\right)   \cr
&\qquad  + x  \left({\frac12\mu s_0 t_n } z^n + v_{n+1} z^{n+1} + \cdots \right) 
+     \left({\frac14 \mu t_n^2} z^{2n} +  w_{2n+1} z^{2n+1} + \cdots  \right) \,.
\ea
\ee
We have achieved our new ansatz for $I_{2 n +1}$.

 As we have seen before, however, with this form of the equation and
no further divisibility assumptions, we are unable to 
make a change of coordinates which would
 put the equation into ``Tate form'' (with monodromy).  If there
is no monodromy, the situation is different and we will return
to that case later.

Now, as a second inductive step, we assume also that $z^{2n+2}$
divides $\Delta$.  This time,
\[ (4u_0w_{2n+1}+4u_1w_{2n}-2v_nv_{n+1})|_{S}\]
must be identically zero 
(since we already know that $(4u_0w_{2n}-v_n^2)|_{S}$  is
identically zero).  We can rewrite that as
\be\label{eq:induct}
 \mu\left(
s_0^2(w_{2n+1}|_{S}) + t_n^2(u_1|_{S})
-s_0t_n(v_{n+1}|_{S})\right) = 0.
\ee

Let $\gamma$ be the greatest common divisor of $s_0|_S$ and $t_n|_S$.
Then any irreducible factor $\phi$
of $(s_0|_S)/\gamma$ must divide $u_1|_S$.
If $\phi$ is not a unit in the coordinate ring of $S$, 
choose an irreducible component
$\Sigma$
of $ \{ \phi =0\}\subset S$.
Then $u_0$ has multiplicity at least two along $\Sigma$ and $u_1$
has multiplicity at least one along $\Sigma$, so
$u=u_0+u_1z+O(z^2)$ has multiplicity at least two along $\Sigma$.  But this
contradicts our hypothesis!

Thus, there are no non-trivial factors of $(s_0|_S)/\gamma$, which implies that
$s_0|_S$ divides $t_n|_S$.  That is, there exists a function $u_n$ such
that
$t_n = -\frac13s_0u_n +O(z)$ (possibly after shrinking $U$).
We can then solve \eqref{eq:induct} by
\be
w_{2n+1} = -\frac13u_nv_{n+1}-\frac19u_1u_n^2 + O(z)\,.
\ee
We replace $t_n$ and $w_{2n+1}$ by $-\frac13s_0u_n$ 
and $-\frac13u_nv_{n+1}-\frac19u_1u_n^2$, respectively, and make the
corresponding adjustments to the other coefficients.

Now when we make a change of coordinates 
\be\label{eq:xshift}
x \mapsto  x+\frac13 u_n z^n  \,,
\ee
the equation takes the form
\be
\label{eq:i2n}
\ba
y^2 = &x^3 +(u_0+u_1z+\cdots+u_{n-1}z^{n-1}+u_nz^n)x^2
+ (\widetilde{v}_{n+1}z^{n+1} + \cdots)x
+ (\widetilde{w}_{2n+2}z^{2n+2} + \cdots)
\ea
\ee
where again $u_0= {\mu s_0^2/4}$ and 
\be
\widetilde{v}_{n+1} = v_{n+1} + {\frac2 3} u_1 u_n \,,\qquad 
\widetilde{w}_{2n+2} = w_{2n+2} + {\frac1 9} u_2 u_n^2 \,.
\ee
(Note that $u_0$, \dots, $u_{n-1}$ are unchanged by this change of
coordinates.)
Thus, we have achieved the same form  of the equation but with $n$ replaced
by $n+1$, i.e., $I_{2n+2}$ ``Tate form" with monodromy, and the inductive step is verified.

\subsection{Case without monodromy}

{
Our induction argument has established that for  $I_m$ with $m\ge10$,
there is always a locally defined change of coordinates which puts the
equation into ``Tate form with monodromy'' when $m$ is even, and into
the form of our new ansatz (which replaces ``Tate form with monodromy''
from \cite{geom-gauge}) when $m$ is odd.

For models without monodromy, there is a further coordinate change which
can be made which puts these equations into ``Tate form without monodromy".
Recall that we detect the lack of monodromy by the condition $\mu \equiv1$.

If $m=2n$ is even, we can apply the coordinate change
\be\label{eq:NoMonoShiftbis}
(x, y) \mapsto \left(x, y+ \frac12 s_0 x\right) \,
\ee
to \eqref{eq:TateI2n} (bearing in mind that $u_0=\frac14\mu s_0^2=\frac14s_0^2$)
to obtain
\begin{equation} 
\ba
y^2+ s_0xy= &x^3 + (u_1z+u_2z^2+ \dots+u_{n-1}z^{n-1}) x^2 
+ (v_n z^n + v_{n+1}z^{n+1} + \dots)x  \cr
&+  (w_{2n} z^{2n} + w_{2n+1}z^{2n+1} + w_{2n+2}z^{2n+2} + \dots) \,.
\ea
\end{equation}
This is ``Tate form without monodromy'' for $I_{2n}$.

On the other hand, if $m=2n+1$ is {{odd}}, we can apply the coordinate change
}
\be\label{eq:ShiftNoMo}
(x, y) \quad \mapsto \left( x\,,\ y+ \frac12 s_0 x 
+ \frac12 t_{n}z^{n}  \right) 
\ee
to \eqref{InMonoNew} to obtain
\be
y^2 +  s_0  x y   +t_{n} z^{n} y
= x^3  +  (u_1 z +\cdots+u_{n-1}z^{n-1})x^2 +  (v_{{n}+1} z^{{n}+1}  + \cdots )x
+ (w_{2{n}+1}  z^{2{n}+1}   + \cdots) \,.
\ee
This is the ``Tate form
without monodromy''
 for $I_{2{n}+1}$.

\subsection{Outliers}

Our inductive proof shows that for $I_m$ with $m\ge10$, there is a 
coordinate change (on a sufficiently small open set $U$) which either puts
the equation into ``Tate form'' or into the form given by our new ansatz.
We also showed this explicitly for $I_m$ with $m\le5$.  What about
the intermediate cases: the cases $I_6$ through $I_9$?

We do not have any general results about these cases to report on here.
However, there are some examples already in the literature which show
that at least sometimes, neither Tate form nor our new ansatz can be
achieved by a coordinate change.  (Although these examples appear in
the literature, it does not seem to have been observed that they
cannot be put into ``Tate form.'')

In 
\cite{Katz:1996xe}, the Cartan deformation of an $E_6$ singularity to $A_5$ 
was shown to be
\be
y^2 =  x^3 + \frac{81 t^4}{64} x^2 - \frac{9 t^2 z^2}{8}x + \frac{z^4}{4}
\, .
\ee
This already satifies our new ansatz for $I_5$, and since there is
no monodromy, it can also be put into ``Tate form'' for $I_5$ by the
coordinate change $(x,y)\mapsto (x,y-\frac98t^2x+\frac12z^2)$, giving
an equation
\be 
y^2-\frac94t^2xy +z^2y=x^3
\, .
\ee
However, this singularity has type $I_6$, so let us attempt to follow the
algorithm in this case.  We have $s_0={-}\frac94t^2$ and $t_2=
1$ 
so at
the next step we would define $u_2 = -3t_2/s_0$ and make the coordinate
change $x\mapsto x+\frac13u_2z^2$.  But since $u_2=\frac43 t^{-2}$,
this cannot be done.

Similarly, for $E_7$ deformed to $A_6$, \cite{Katz:1996xe} found
\begin{align}
y^2 = & \widetilde{x}^3 + \left( 729 t^6 + 63t^2 z\right)\widetilde{x}^2 
+ \left(13122 t^8 z + 243 t^4 z^2 - 16 z^3\right)\widetilde{x} \nonumber\\
      & + \left( 59049t^{10} z^2 - 2187t^6 z^3 \right)
      \, .
\end{align}
The first steps in Tate's algorithm are accomplished by the
coordinate change $\widetilde{x}=x-9t^2z$, which leaves us with the 
equation
\be
y^2=x^3+\left( 729 t^6 + 36t^2 z\right)x^2
+ \left(-648 t^4 z^2 - 16 z^3\right)x
+144 z^4t^2
\, .
\ee
Again, this satisfies our new ansatz for $I_5$, although since there is
no monodromy,
it can also be put into ``Tate form'' for $I_5$ by the
coordinate change: $(x,y)\mapsto (x,y-27t^3x+12tz^2)$, giving
an equation
\be
y^2 -  54 t^3 \, xy  +  24 t z^2 \,y = x^3 + 36tz\,x^2  - 16z^3\, x   
\, .
\ee
This singularity has type $I_7$, so we again attempt to follow the
algorithm.  We have $s_0={-}54t^3$ and $t_2=24t$, so that 
$u_2=-3t_2/s_0=\frac43t^{-2}$
is ill-defined.  As in the previous case, this obstructs us from 
carrying out the algorithm to put the equation into $I_6$ or $I_7$ form.


\subsection{Summary for $I_m$}

In summary, for the $I_m$ case we have shown that without further assumptions, the following forms for the elliptic fibration can be achieved: 
\bigskip

\begin{tabular}{|l|l|}
\hline
Type & Form\cr \hline\hline
 $I_2$, $I_4$, $I_{2n}$, $n>5$ 
with/without monodromy
 & ``Tate form" can be achieved.\cr\hline
 $I_3$, $I_5$, $I_{2n+1}$, $n>5$ with monodromy & ``Tate form" not possible near zeros of $s_0$ in $S$. \cr
 									       & New ansatz (\ref{InMonoNew}) can be achieved. \cr\hline
 $I_3$, $I_5$, $I_{2n+1}$, $n>5$ without monodromy&	``Tate form" can be achieved.	 \cr \hline
 $I_6$, $I_7$, $I_8$, $I_9$ :& Neither ``Tate form" nor new ansatz can be achieved.\cr\hline
 \end{tabular}

\bigskip


\section{The $II$, $III$, and $IV$ cases}


For Kodaira fibers of types $II$, $III$, and $IV$, we revert to Weierstrass
form as our starting point.  In these cases the Kodaira criterion is
very straightforward, and can be applied immediately.

\begin{enumerate}
\item To obtain Kodaira type $II$ (which has no enhanced  gauge symmetry),
we need $z \ |\ f_0$ and $z\ |\ g_0$.  We may absorb these coefficients into
$f_1$ and $g_1$, respectively, and find an equation of the form
\be
y^2 = x^3 + (f_1z + f_2z^2 + \cdots) x + (g_1z + g_2z^2+\cdots).
\ee
This is ``Tate form'' for type $II$, and has type $II$ provided that
$z \not| g_1$.
\item To obtain Kodaira type $III$ (which has $SU(2)$
local  gauge symmetry),
we need in addition
$z\ |\ g_1$.  We may thus absorb $g_1$ into $g_2$ 
and find an equation of the form
\be
y^2 = x^3 + (f_1z + f_2z^2 + \cdots) x + ( g_2z^2+g_3z^3+\cdots).
\ee
This is ``Tate form'' for type $III$, and has type $III$ provided that
$z \not| f_1$.
\item To obtain Kodaira type $IV$ (which has either $SU(3)$
or $Sp(1)$
local  gauge symmetry),
we need in addition
$z\ |\ f_1$.  We may thus absorb $f_1$ into $f_2$ 
and find an equation of the form
\be
y^2 = x^3 + (f_2z^2 + f_3z^3 + \cdots) x + ( g_2z^2+g_3z^3+\cdots).
\ee
This is ``Tate form'' for type $IV$, and has type $IV$ provided that
$z \not| g_2$.

The gauge symmetry is determined by the monodromy, which according to
Tate \cite{MR0393039}, depends on whether or not $g_2|_S$ is a square.
If $g_2|_S$ is not a square, then there is monodromy and the local
gauge symmetry
is $Sp(1)$.  If $g_2|_S$ is a square, then there is no
monodromy and the local gauge symmetry is $SU(3)$.
\end{enumerate}

\section{The $I_m^*$ case}

For fibers of type $I_m^*$, we once again start in Weierstrass form.
Kodaira tells us that to have type $I_0^*$, we must have
$z^2\ |\ f$, $z^3\ |\ g$, and the order of $\Delta$ along $S$ must be
exactly $6$.  
\be
 y^2 = x^3 + (f_2z^2+f_3z^3+\cdots)x + (g_3z^3+g_4z^4+\cdots)
\ee
This is ``Tate form'' for type $I_0^*$.
The monodromy in this case is quite subtle, but involves analyzing the
branching behavior of the cubic polynomial
\be
x^3 + \frac{f}{z^2} x + \frac{g}{z^3}.
\ee

\subsection{Step 1}

For $I_m^*$ with $m>0$, the condition is slightly different: the
orders of $f$, $g$, and $\Delta$ must be exactly $2$, $3$ and $m+6$.
We write the leading terms in the discriminant as
\be
\Delta = (4f_2^3+27g_3^2)z^6 + O(z^7).
\ee
and note that the first term must vanish whenever $z^7\ |\ \Delta$.
In this case,
by an argument in appendix~\ref{app:ufd}, there exists a function $u_1$ (possibly
after shrinking $U$) such that 
\be
f_2 = -\frac13 u_1^2  +O(z)\,, \qquad
g_3 = \frac2{27}u_1^3 + O(z)\,.
\ee
We replace $f_2$ by ${ -\frac13 u_1^2}$  and $g_3$ by $\frac2{27}u_1^3$,
modifying the higher coefficients as necessary.
Now following Tate, we make a change of coordinates
\be
(x,y) \mapsto \left(x+\frac13u_1z,y\right) \,,
\ee
which transforms the defining equation to
\be
y^2 = x^3+u_1zx^2 
 +(f_3z^3+f_4z^4+\cdots)x+
(g_4+\frac13u_1f_3)z^4+
(g_5+\frac13u_1f_4)z^5+\cdots \,.
\ee
This is ``Tate form'' for type $I^*_1$.
To simplify later formulas, we set $\widetilde g_j = g_j + \frac{1}{3}u_1 f_{j-1}$, and write the equation in the form
\be \label{eq:tateI1star}
y^2 = x^3+u_1zx^2 
 +(f_3z^3+f_4z^4+\cdots)x+
\widetilde g_4z^4+
\widetilde g_5 z^5+\cdots \,.
\ee

\subsection{Step 2}

At the next order the discriminant is
\be
\Delta = 4 u_1^3 \widetilde g_4 z^7 + O(z^8)
\, .
\ee
The condition that the discriminant vanishes to the next order is
\be
\widetilde g_4 = 0
\, .
\ee
The fibration then takes the form
\be
y^2 = x^3 + u_1 z x^2 + \left( f_3 z^3 + f_4 z^4 + \cdots \right) x + \left( \widetilde g_5 z^5 + \widetilde g_6 z^6 + \cdots \right)
\, .
\ee
This is ``Tate form'' for $I_2^*$.

\subsection{Step 3}

At the next order the discriminant is
\be
\Delta = {{u_1^2  \left( 4 u_1 \widetilde g_5 - f_3^2  \right)}}  z^8 + O(z^9)
\, .
\ee
Analogously to the case  $I_n$, we can find functions $s_0$ and $\mu_1$ such that $\mu_1 |_S$ is square-free and a function $t_2$ such that
\be
u_1 = \frac{1}{4} \mu_1 s_0^2 \, , \qquad  f_3 = \frac{1}{2} \mu_1 s_0 t_2
\,. 
\ee
{{
We can then solve for $\widetilde g_5$ as
\be
\widetilde g_5 =  \frac{1}{4} \mu_1 t_2^2 
\ee
giving a fibration of the form
\be
y^2 = x^3 + \frac{1}{4} \mu_1 s_0^2 z x^2 + \left(\frac{1}{2} \mu_1 s_0 t_2 z^3 +  f_4 z^4 + \cdots \right) x + \left( \frac{1}{4} \mu_1 t_2^2 z^5 + \widetilde g_6 z^6 + \cdots \right)\,.
\ee
}}
The necessary coordinate change to put this in Tate form for type $I_{3}^*$,
\be
(x,y) \rightarrow \left(x-{\frac{{t_2}z^2}{ s_0}},y\right) 
\ee
does not exist near zeros of $s_0$ along $S$.

\subsection{Induction}

We now set up an induction which assumes that the orders of vanishing of
$f$ and $g$ are precisely $2$ and $3$, respectively, and that $z^{k+6}\ |\ \Delta$
for some $k < m$.  Our induction will assume a certain form for the
equation (to be described shortly), and proceed to derive the corresponding
form for $k+1$.  Our equations will all have the general form
\be \label{eq:instargeneral}
y^2 = x^3+u x^2 + vx + w
\ee
with $z\ |\ u$,  $z^3\ |\ v$ and $z^4\ |\ w$.  We let $u=\widehat{u}z.$ 
  Note that we cannot have
$z\ |\ \widehat{u}$, or else we would be in a different branch of Kodaira's
classification (as in that case $f$ and $g$ would vanish to order greater
than $2$ and $3$, respectively).

As an initial hypothesis we assume that $m\ge4$ so that $\Delta$ vanishes
to order at least $10$.  The reason is similar to the $I_m$ case: under
this hypothesis, $\widehat{u}|_S$ can have no zeros.  For if there were any
zero of $\widehat{u}|_S$ then the multiplicities of $f$, $g$, and $\Delta$
at such a point would be at least $(3,4,11)$, and that forces the point
into the ``non-minimal'' part of Kodaira's classification, with the
consequent massless tensors, light strings, etc.

Now to our induction.  
We start by assuming that we have achieved
``Tate form'' for some $I_k^*$ 
with $k<m$ and will show how to increase $k$.
If $k=2n-1$ is odd, we assume by inductive
hypothesis that the expansion of $u=\widehat{u}z=u_1z+u_2z^2+\cdots+u_nz^n$ 
has no term divisible by $z^{n+1}$ and that 
the  ``Tate form'' for $I^*_{2n-1}$ holds (ignoring the monodromy
condition): namely, that
 $z^{n+2} \ |\ v$ and $z^{2n+2}\ |\ w$.  Note
that the form achieved in \eqref{eq:tateI1star} is exactly of this
type for $n=1$, under the simple assumption that $z^7\ |\ \Delta$.
Thus, our induction has a place to begin.

Under this assumption, there is only one contribution to the leading
term in the discriminant \eqref{eq:terms}:
\be
\Delta = 4u_1^3w_{2n+2} z^{2n+5} + O(z^{2n+6}).
\ee
Since we are assuming that $k<m$,
the leading term must vanish, that is,
$z\ |\ w_{2n+2}$ or equivalently $z^{2n+3}\ |\ w$ (since 
 $\widehat{u}|_S$ is not 
identically zero).  We can thus absorb $w_{2n+2}$ into $w_{2n+3}$ by
{{adjusting}} the latter, after which  we have achieved the 
conditions
$z^{n+2}\ |\ v$ and $z^{2n+3}\ |\ w$ when $z^{2n+6}\ |\ \Delta$.
This we will take to be our corresponding inductive hypothesis when
$k=2n$ is even.  (This is ``Tate form'' for $I^*_{2n}$, ignoring
the monodromy condition.)

As the second step in the induction, we now assume we are in that form.
This time, the leading contribution to the discriminant contains two
terms:
\be
\Delta = u_1^2(4u_1w_{2n+3}-v_{n+2}^2) z^{2n+6} +O(z^{2n+7}).
\ee
Since again $k<m$ by our inductive assumption, we must have
\be \label{eq:Instardisc}
(4u_1w_{2n+3}-v_{n+2}^2)|_S \equiv 0.
\ee
Since $u_1|_S$ has no zeros by our initial hypothesis, 
we may find a function $u_{n+1}$ 
(possibly after shrinking $U$) such that
\be
v_{n+2} = -\frac23u_1u_{n+1} +O(z).
\ee
Then in order to have the 
vanishing specified
 in \eqref{eq:Instardisc} we must
also have
\be
w_{2n+3} = \frac19u_1u_{n+1}^2 +O(z).
\ee
Replace $v_{n+2}$ by $-\frac23u_1u_{n+1}$ and $w_{2n+3}$ by
$\frac19u_1u_{n+1}^2$, and adjust the other coefficients
accordingly.  Now we can make a change of coordinates
$(x,y)\mapsto(x+\frac13 u_{n+1}z^{n+1},y)$.
This 
adds $u_{n+1}z^{n+1}$ to the coefficient of $x^2$, and increases
the order of vanishing of $v$ and $w$ by one each.  That is, we have
$z^{2n+7}\ |\ \Delta$ while $z^{n+3}\ |\ v$ and
$z^{2n+4}\ |\ w$.  This reproduces our inductive hypotheses for
$k=2n+1$, so our induction argument is complete.

\bigskip

The final remark is about the monodromy in $I_m^*$ cases.  According
to Tate's original algorithm, for $I^*_{2n-1}$ the test for monodromy is
whether $w_{2n+2}|_S$ has a square root.  (This distinguishes between
$SO(4n+5)$ when there is no square root, and
$SO(4n+6)$ when there is a square root.)
Similarly, for $I^*_{2n}$ the test for monodromy is whether
$(4u_1w_{2n+3}-v_{n+2}^2)|_S$ has a square root.  (This
distinguishes between
$SO(4n+7)$ when there is no square root, and
$SO(4n+8)$ when there is a square root.)

\subsection{Summary of the $I_m^*$ case}

We have thus shown that we can always write a Weierstrass fibration in ``Tate form'' for types $I^*_1$,$I^*_2$, and $I^*_m$ for $m\geq 4$. 
For the case $I^*_3$, where ``Tate form'' is not always achievable, we can find simple examples in the literature that exhibit this behaviour. 
In particular \cite{anomalies}, 
the unfolding of $E_8$ to $D_7$ is described by the fibration
\be
y^2 = x^3 +  t^2zx^2 +2tz^3x+ z^5
\, .
\ee
This is already in ``Tate form'' for $I_2^*$ with $u_1=t^2$, $v_3=2t$,
and $w_5=1$.  However, the generic singularity is of type $I_3^*$; if
we attempt to follow the algorithm, then the next coordinate change
involves $u_3=-3v_3/2u_1=-3t^{-1}$ which is ill-defined.  Thus, this
example cannot be put into ``Tate form'' for $I_3^*$.

\section{The $IV^*$, $III^*$, and $II^*$ cases}  %

For Kodaira fibers of types $IV^*$, $III^*$, and $II^*$, we again
revert to Weierstrass
form as our starting point.  In these cases the Kodaira criterion is
very straightforward, and can be applied immediately.

\begin{enumerate}
\item To obtain Kodaira type $IV^*$ (which has either $E_6$
or $F_4$ local  gauge symmetry),
we need $z^3 \ |\ f$ and $z^4\ |\ g$.  We may thus choose our expansion
to take the form
\be
y^2 = x^3 + (f_3z^3 + f_4z^4 + \cdots) x + (g_4z^4 + g_5z^5+\cdots).
\ee
This is ``Tate form'' for type $IV^*$, and has type $IV^*$ provided that
$z \not| g_4$.
There is also a monodromy question in this case, measured by whether
$g_4|_S$ is a square or not.  If $g_4|_S$ is not a square, there is monodromy
and the local gauge symmetry is $F_4$.  If $g_4|_S$ is a square,
there is no monodromy and the local gauge symmetry is $E_6$.
\item To obtain Kodaira type $III^*$ (which has $E_7$
local  gauge symmetry),
we need in addition
$z\ |\ g_4$.  We may thus absorb $g_4$ into $g_5$ 
and find an equation of the form
\be
y^2 = x^3 + (f_3z^3 + f_4z^4 + \cdots) x + ( g_5z^5+g_6z^6+\cdots).
\ee
This is ``Tate form'' for type $III^*$, and has type $III^*$ provided that
$z \not| f_3$.
\item To obtain Kodaira type $II^*$ (which has $E_8$
local  gauge symmetry),
we need in addition
$z\ |\ f_3$.  We may thus absorb $f_3$ into $f_4$ 
and find an equation of the form
\be
y^2 = x^3 + (f_4z^4 + f_5z^5 + \cdots) x + ( g_5z^5+g_6z^6+\cdots).
\ee
This is ``Tate form'' for type $II^*$, and has type $II^*$ provided that
$z \not| g_5$.
\item
Finally, if $z\ |\ g_5$ we have reached the ``non-minimal'' line on
Kodaira's table; algorithmically, we should alter our Weierstrass model
to one of lower degree and repeat the algorithm.
\end{enumerate}


\section{Conclusions}

In this paper we re-analyzed the ``Tate forms" that were introduced in \cite{geom-gauge} in the light
of their general validity and found that, except in a few instances, the Tate algorithm can be carried through without many modifications. 
Specifically we found that, except for  $I_6$, $I_7$, $I_8$, $I_9$ and $I_3^*$, the ``Tate forms" 
can be achieved for the cases $I_{n}$ without monodromy and for $I_{m}^*$. 
For $I_{2n+1}$ with monodromy we can generically (i.e. without making any further assumptions about
divisibility of the sections) only achieve the new form (\ref{InMonoNew}). The main obstruction stems from changes of variables that are 
necessary in the algorithm, but which may involve poles on the divisor over which the singularity resides. 

Furthermore, we demonstrated with a simple example that Tate forms may not hold globally over the entire
base. For the example in section 3 with an $I_2$ singularity
and in appendix C with $SU(5)$ there are global obstructions to achieving Tate form. For the case of $SU(5)$ we only 
provided one possible recipe for constructing a globally-obstructed Tate form. 
It would be interesting to find explicit geometries realizing these (or analogous) criteria. 
These geometries could, in particular, be relevant for GUT model building 
with $SU(5)$ gauge group by giving more freedom in their construction.


\section*{Acknowledgements}

We thank A. Grassi and W. Taylor for discussions. 
SK and DRM thank the Simons Workshop in Mathematics and Physics for
hospitality at the inception of this project, and IPMU, University of
Tokyo, for hospitality at a later stage.
SSN thanks the Caltech theory group for their generous hospitality,
and JS thanks IPMU, University of Tokyo, for hospitality.
This research was partially supported by the
National Science
Foundation under grants DMS-1007414, DMS-05-55678,
and PHY05-51164, 
and by World Premier International 
Research Center Initiative (WPI Initiative), MEXT, Japan.


\appendix

\section[``Tate forms'' from {[4]}]{``Tate forms'' from \cite{geom-gauge}} \label{app:tateform}

We start with an equation in the general form
\be \label{eq:tatemain}
y^2+a_1xy+a_3y = x^3+a_2x^2+a_4x+a_6.
\ee
(In the main body of this paper, $s$ and $t$ were used on the left side
of the equation in place of $a_1$ and $a_3$, while $u$, $v$ and $w$
were used on the right side of the equation in place of $a_2$,
$a_4$, and $a_6$.)
Table~\ref{tab:tatealg}, which is reproduced from the first part of Table~2
in \cite{geom-gauge}, gives special forms of \eqref{eq:tatemain} which
lead to enhanced gauge symmetry---these have come to be called ``Tate forms,''
but, as emphasized in the body of this paper, they serve as convenient
ans\"atze which do not always apply.  One piece of notation in this
Table needs explanation: a superscript of $^s$ on the Kodaira symbol
indicates no monodromy, while a superscript of $^{ns}$ or $^{ss}$ indicates
monodromy.  (For $I_0^*$, the only case in which $^{ss}$ appears, there
are two types of monodromy and the notation distinguishes between them.)

We modified Table~2 of \cite{geom-gauge} by changing $k$ to $n$ to match
the notation of this paper, and by correcting and completing the ``group''
column in the table, 
according to the more precise conclusions about gauge groups
which were found some years later in \cite{LieF}.

\begin{table}[ht]
{\footnotesize
\begin{center}
\begin{tabular}{|c|c|c|c|c|c|c|c|} \hline
 type & group & $ a_1$ &
$a_2$ & $a_3$ &$ a_4 $& $ a_6$ &$\Delta$ \\ \hline $I_0 $ & --- &$ 0 $ &$ 0
$ &$ 0 $ &$ 0 $ &$ 0$ &$0$ \\ \hline $I_1 $ & --- &$0 $ &$ 0 $ &$ 1 $ &$ 1
$ &$ 1 $ &$1$ \\ \hline $I_2 $ &$SU(2)$ &$ 0 $ &$ 0 $ &$ 1 $ &$ 1 $ &$2$ &$
2 $ \\ \hline $I_{3}^{ns} $ & $Sp(1)$ &$0$ &$0$ &$2$ &$2$ &$3$ &$3$ \\ \hline
$I_{3}^{s}$ & $SU(3)$ &$0$ &$1$ &$1$ &$2$ &$3$ &$3$ \\ \hline
$I_{2n}^{ns}$ &$ Sp(n)$ &$0$ &$0$ &$n$ &$n$ &$2n$ &$2n$ \\ \hline
$I_{2n}^{s}$ &$SU(2n)$ &$0$ &$1$ &$n$ &$n$ &$2n$ &$2n$ \\ \hline
$I_{2n+1}^{ns}$ &$Sp(n)$ & $0$ &$0$ &$n+1$ &$n+1$ &$2n+1$ &$2n+1$
\\ \hline $I_{2n+1}^s$ &$SU(2n+1)$ &$0$ &$1$ &$n$ &$n+1$ &$2n+1$ &$2n+1$
\\ \hline $II$ & --- &$1$ &$1$ &$1$ &$1$ &$1$ &$2$ \\ \hline $III$ &$SU(2)$ &$1$
&$1$ &$1$ &$1$ &$2$ &$3$ \\ \hline $IV^{ns} $ &$Sp(1)$ &$1$ &$1$ &$1$
&$2$ &$2$ &$4$ \\ \hline $IV^{s}$ &$SU(3)$ &$1$ &$1$ &$1$ &$2$ &$3$ &$4$
\\ \hline $I_0^{*\,ns} $ &$G_2$ &$1$ &$1$ &$2$ &$2$ &$3$ &$6$ \\ \hline
$I_0^{*\,ss}$ &$SO(7)$ &$1$ &$1$ &$2$ &$2$ &$4$ &$6$ \\ \hline $I_0^{*\,s}
$ &$SO(8)^*$ &$1$ &$1$ &$2$ &$2$ &$4$ & $6$ \\ \hline $I_{1}^{*\,ns}$
&$SO(9)$ &$1$ &$1$ &$2$ &$3$ &$4$ &$7$ \\ \hline $I_{1}^{*\,s}$ &$SO(10) $
&$1$ &$1$ &$2$ &$3$ &$5$ &$7$ \\ \hline $I_{2}^{*\,ns}$ &$SO(11)$ &$1$ &$1$
&$3$ &$3$ &$5$ &$8$ \\ \hline $I_{2}^{*\,s}$ &$SO(12)^*$ &$1$ &$1$ &$3$
&$3$ &$5$&$8$\\ \hline 
$I_{2n-3}^{*\,ns}$ &$SO(4n+1)$ &$1$ &$1$ &$n$ &$n+1$
&$2n$ &$2n+3$ \\ \hline $I_{2n-3}^{*\,s}$ &$SO(4n+2)$ &$1$ &$1$ &$n$ &$n+1$
&$2n+1$ &$2n+3$ \\ \hline $I_{2n-2}^{*\,ns}$ &$SO(4n+3)$ &$1$ &$1$ &$n+1$
&$n+1$ &$2n+1$ &$2n+4$ \\ \hline $I_{2n-2}^{*\,s}$ &$SO(4n+4)^*$ &$1$ &$1$
&$n+1$ &$n+1$ &$2n+1$ 
&$2n+4$ \\ \hline $IV^{*\,ns}$ &$F_4 $ &$1$ &$2$ &$2$ &$3$ &$4$
&$8$\\ \hline $IV^{*\,s} $ &$E_6$ &$1$ &$2$ &$2$ &$3$ &$5$ & $8$\\ \hline
$III^{*} $ &$E_7$ &$1$ &$2$ &$3$ &$3$ &$5$ & $9$\\ \hline $II^{*} $
&$E_8\,$ &$1$ &$2$ &$3$ &$4$ &$5$ & $10$ \\ \hline
 non-min & --- &$ 1$ &$2$ &$3$ &$4$ &$6$ &$12$ \\ \hline
\end{tabular}
\caption{``Tate forms'' (from \cite{geom-gauge})} \label{tab:tatealg}
\end{center}
}
\end{table}

The special forms are specified by declaring various coefficients
$a_i$ to vanish along the discriminant  component
$\{z=0\}$ to various
orders (at a minimum): the orders are specified in Table~\ref{tab:tatealg}.
By finding the lowest row in the table for which the vanishing conditions
are satisfied, we determine the Kodaira fiber type and monodromy.
In addition, the asterisks next to $SO(4n+4)$ indicate that one more
condition must be fulfilled in those cases: for $SO(8)$ we must have that
\[\left. \frac{a_2^2-4a_4}{z^4}\right|_{z=0}\]
is a square, whereas for $SO(4n+4)$ with $n\ge3$ we need that
\[\left. \frac{a_4^2-4a_2a_6}{z^{2k+2}}\right|_{z=0}\]
is a square.

Now in general, to pass from \eqref{eq:tatemain} to Weierstrass form involves
completing the square of the left hand side of \eqref{eq:tatemain},
and then completing the cube on the right hand side.  For the forms specified
in Table~\ref{tab:tatealg}, usually part of this completing the square and/or
cube can be done without disturbing the vanishing conditions.  This gives
a more compact version of the ``Tate form'' in each case, in which some
of the coefficients in \eqref{eq:tatemain} are suppressed altogether.
The results of this operation are displayed in Table~\ref{tab:tatealgcompact},
in which an entry ``$\infty$'' indicates that a coefficient is to be set
to zero.  The same extra condition for $SO(4n+4)$ must be applied
as in the original form.

\begin{table}[ht]
{\footnotesize
\begin{center}
\begin{tabular}{|c|c|c|c|c|c|c|c|} \hline
type & group & $ a_1$ & $a_2$ & $a_3$ &$ a_4 $& $ a_6$ &$\Delta$ \\ \hline
$I_0 $ & --- &$\infty$ &$ \infty$ &$\infty$ &$ 0 $ &$ 0$ &$0$ \\ \hline
$I_1 $ & --- &$\infty$ &$ 0 $ &$\infty$ &$ 1$ &$ 1 $ &$1$ \\ \hline
$I_2 $ &$SU(2)$ &$\infty$ &$ 0 $ &$\infty$ &$ 1 $ &$2$ &$2 $ \\ \hline
$I_{3}^{ns} $ & $Sp(1)$ &$\infty$ &$0$ &$\infty$ &$2$ &$3$ &$3$ \\ \hline
$I_{3}^{s}$ &$SU(3)$ &$0$ &$\infty$ &$1$ &$2$ &$3$ &$3$ \\ \hline
$I_{2n}^{ns}$ &$ Sp(n)$ &$\infty$ &$0$ &$\infty$ &$n$ &$2n$ &$2n$ \\ \hline
$I_{2n}^{s}$ &$SU(2n)$ &$0$ &$1$ &$\infty$ &$n$ &$2n$ &$2n$ \\ \hline
$I_{2n+1}^{ns}$ &$Sp(n)$ & $\infty$ &$0$ &$\infty$ &$n+1$ &$2n+1$ &$2n+1$\\ \hline
$I_{2n+1}^s$ &$SU(2n+1)$ &$0$ &$1$ &$n$ &$n+1$ &$2n+1$ &$2n+1$\\ \hline
$II$ & --- &$\infty$ &$\infty$ &$\infty$ &$1$ &$1$ &$2$ \\ \hline
$III$ &$SU(2)$ &$\infty$&$\infty$ &$\infty$ &$1$ &$2$ &$3$ \\ \hline
$IV^{ns} $ &$Sp(1)$ &$\infty$ &$\infty$ &$\infty$&$2$ &$2$ &$4$ \\ \hline
$IV^{s}$ &$SU(3)$ &$\infty$ &$\infty$ &$1$ &$2$ &$3$ &$4$\\ \hline
$I_0^{*\,ns} $ &$G_2$ &$\infty$ &$\infty$ &$\infty$ &$2$ &$3$ &$6$ \\ \hline
$I_0^{*\,ss}$ &$SO(7)$ &$\infty$ &$1$ &$\infty$ &$2$ &$4$ &$6$ \\ \hline
$I_0^{*\,s}$ &$SO(8)^*$ &$\infty$ &$1$ &$\infty$ &$2$ &$4$ & $6$ \\ \hline
$I_{1}^{*\,ns}$&$SO(9)$ &$\infty$ &$1$ &$\infty$ &$3$ &$4$ &$7$ \\ \hline
$I_{1}^{*\,s}$ &$SO(10) $&$\infty$ &$1$ &$2$ &$3$ &$5$ &$7$ \\ \hline
$I_{2}^{*\,ns}$ &$SO(11)$ &$\infty$ &$1$&$\infty$ &$3$ &$5$ &$8$ \\ \hline
$I_{2}^{*\,s}$ &$SO(12)^*$ &$\infty$ &$1$ &$\infty$&$3$ &$5$&$8$\\ \hline 
$I_{2n-3}^{*\,ns}$ &$SO(4n+1)$ &$\infty$ &$1$ &$\infty$ &$n+1$&$2n$ &$2n+3$ \\ \hline
$I_{2n-3}^{*\,s}$ &$SO(4n+2)$ &$\infty$ &$1$ &$n$ &$n+1$&$2n+1$ &$2n+3$ \\ \hline
$I_{2n-2}^{*\,ns}$ &$SO(4n+3)$ &$\infty$ &$1$ &$\infty$&$n+1$ &$2n+1$ &$2n+4$ \\ \hline
$I_{2n-2}^{*\,s}$ &$SO(4n+4)^*$ &$\infty$ &$1$&$\infty$ &$n+1$ &$2n+1$ &$2n+4$\\ \hline
$IV^{*\,ns}$ &$F_4 $ &$\infty$ &$\infty$ &$\infty$ &$3$ &$4$&$8$\\ \hline
$IV^{*\,s} $ &$E_6$ &$\infty$ &$\infty$ &$2$ &$3$ &$5$ & $8$\\ \hline
$III^{*} $ &$E_7$ &$\infty$ &$\infty$ &$\infty$ &$3$ &$5$ & $9$\\ \hline
$II^{*} $&$E_8\,$ &$\infty$ &$\infty$ &$\infty$ &$4$ &$5$ & $10$ \\ \hline
non-min & --- &$\infty$ &$\infty$ &$\infty$ &$4$ &$6$ &$12$ \\ \hline
\end{tabular}
\caption{``Tate forms'' (compact version)} \label{tab:tatealgcompact}
\end{center}
}
\end{table}

As mentioned earlier, 
it is these compact versions of Tate forms which we have referred to repeatedly
in the body of the paper.


\section{Lemmas using unique factorization} \label{app:ufd}

In this appendix we prove the lemmas that were used in the text in 
implementing Tate's algorithm.  

We keep the notation in the main text: $B$ is the smooth base, $S$ a smooth
divisor over which enhancement occurs, and $U\subset B$ an affine open set.
By smoothness, the
rings of algebraic functions on $U$ or $U\cap S$ are unique factorization
domains.  In these UFDs, the units are just the nowhere vanishing functions.
 
Recall that we are identifying the restricted leading coefficients 
$f_0|_S$ and $g_0|_S$ with
well-defined as functions on $S\cap U$.  (More generally, they
will be well-defined sections of line bundles on $S$.)

We use the notation $X=Y+O(z)$ to indicate that $X-Y\in \mathcal{I}_S$, 
since $z$ is a local defining equation for $S$.

We will routinely extend functions on $U\cap S$ to functions on $U$.  If
we were only dealing with regular functions, this would be automatic
since the coordinate ring of $U\cap S$ is a quotient of the ring of
regular functions on $U$.  In the more general situation 
of algebraic functions, we may have to shrink $U$ to keep the functions
single-valued.  

\begin{lemma}
If $(4f_0^3+27g_0^2)|_S=0$ then possibly after 
shrinking
$U$, there exists a function $u_0$ on $U$ such that $f_0|_S=-\frac13u_0^2|_S$
and $g_0|_S=\frac2{27}u_0^3|_S$, i.e., 
$f_0=-\frac13u_0^2+O(z)$ and $g_0=\frac2{27}u_0^3+O(z)$.
\end{lemma}

\bigskip\noindent
{\em Proof:\/} We factor the restrictions of $f_0$ and $g_0$ into
irreducibles

\begin{equation}
f_0|_S = \prod_{i=1}^m f_i^{\alpha_i},\qquad g_0|_S=
\prod_{j=1}^n g_j^{\beta_j},
\end{equation}
unique up to ordering and multiplication by units.  From 
$(4f_0^3+27g_0^2)|_S=0$ and unique factorization, we see that $m=n$, and that
after reordering the $f_i$ and $g_j$ if necessary that the $f_i$ and $g_i$
are equal up to multiplication by a unit.  We conclude
that there are integers $\gamma_i$
such that for all $i$ we have $\alpha_i=2\gamma_i$ and $\beta_i=3\gamma_i$.

We put 
\begin{equation}
v_0=c\prod_{i=1}^m f_i^{\gamma_i}
\end{equation}
for a constant $c$ and 
demand that $f_0|_S = -\frac13v_0^2$ and $g_0|_S=\frac2{27}v_0^3$.  The
first condition requires $c^2=-3$ and the second condition fixes the choice
of the square root to determine $c$.  
We now let $u_0$ be any function on $U$ restricting to
$v_0$ on $S$ (shrinking $U$ if necessary) and we are done.

\begin{lemma}
Given a function $u_0$ whose restriction to $S$
is not identically zero, then possibly after shrinking $U$
there exist functions $s_0$ and $\mu$ such that $\mu|_S$ is square-free,
and $u_0=\frac14\mu s_0^2+O(z)$.
\end{lemma}

\bigskip\noindent
{\em Proof:\/} We write
\begin{equation}
u_0|_S = \prod_{i=1}^m u_i^{\alpha_i}
\end{equation}
Without loss of generality we may suppose that $\alpha_i$ is odd for $i\le k$
and even for $k+1\le i\le m$.  We put $\alpha_i=2\beta_i+1$ for $i\le k$ and
$\alpha_i=2\beta_i$ for $k+1\le i\le m$.
Then we put
\begin{equation}
t_0=\prod_{i=1}^m u_i^{\beta_i}\qquad \nu=4\prod_{i=1}^k u_i,
\end{equation}
so that $\nu$ is square-free and $u_0|_S=\frac14\nu t_0^2$.

Now let $s_0$ and $\mu$ be any functions on $U$ restricting to $t_0$ and $\nu$
respectively on $S$ (shrinking $U$ if necessary), 
so that $u_0=\frac14\mu s_0^2+O(z)$ and we are done.

\begin{lemma}
If $(4\mu s_0^2\widetilde{g}_2 - f_1^2)|_S=0$ 
and $\mu|_S$ is 
square-free, then possibly after shrinking $U$ there exists a function
$t_1$ such that $f_1=\frac12\mu s_0t_1+O(z)$.
\end{lemma}

\bigskip\noindent
{\em Proof:\/} Since $\mu|_S$ is square-free, we have a factorization
\begin{equation}
\mu|_S=\prod_{i=1}^n \mu_i
\end{equation}
with distinct factors.  We also factor
\begin{equation}
\widetilde{g}_2|_S = \prod_{j=1}^m h_j^{\beta_j}
\end{equation}
Since the exponents in $(4\mu s_0^2\widetilde{g}_2)|_S$ are even, we
can reorder the $h_j$ if necessary to achieve $h_j=\mu_j$ up to a unit
and $\beta_j$ odd
for $j\le n$, and $\beta_j$ even for $j>n$.  As in the proofs of the earlier
lemmas, we can easily write down a function $s_1$ on $U\cap S$ such
that 
\begin{equation}
f_1|_S=\frac12\left(\mu s_0\right)|_Ss_1.
\end{equation}
We then extend $s_1$ to a function $t_1$ on $U$ (shrinking $U$ if necessary)
and we are done.

\begin{lemma} \label{lem:4}
If 
\be \label{eq:lemma4}
(s_0^3\widetilde{g}_3 - s_0^2t_1f_2 - t
   _1^3)|_S \, ,
\ee
then possibly after shrinking $U$, there exists a function $u_1$ on $U$ such
that $t_1=-\frac13 s_0u_1+O(z)$.
\end{lemma}

\bigskip\noindent
{\em Proof:\/} We argue that $s_0|_S$ divides $t_1|_S$
by induction on the number of irreducible factors of $s_0|_S$.  If
there are none, we are done.  If there is an irreducible factor $\alpha$,
then it divides the first two terms in eq.~\eqref{eq:lemma4} so it must
divide $t_1^3|_S$ and hence $t_1|_S$.  This implies that $\alpha^3$ divides
all three terms in eq.~\eqref{eq:lemma4}; dividing by $\alpha^3$ reduces the
number of irreducible factors of $s_0|_S$ and by induction we are finished.


\section{Global obstructions for $SU(5)$}\label{app:C}

In section \ref{sec:obstruction} we gave an explicit example of an $I_2$ Weierstrass fibration that could not be written globally in Tate-type normal form. Although it seems clear that the same type of obstruction can appear at higher order, it would be nice to have a analogous example for $SU(5)$. Unfortunately it is not easy to mechanically construct such an example. Instead, we simply give a list of sufficient criteria for such an example to exist. Note that we have not proven that it is possible to satisfy all of the following conditions simultaneously.

Consider a base $B$ with an effective anti-canonical divisor $-K$ and an effective divisor ${S}$ on $B$ such that $-12K - 5{S}$ is effective. As the discriminant $\Delta$ is a section of $-12K$, this allows us to build a fibration that vanishes to fifth order on ${S}$. In the following, we will make frequent use of the exact sequence of sheaves
\begin{align}
0 \rightarrow \mathcal{O}_{B}\left(-nK -(m+1){S}) \right)
\stackrel{\cdot z}{\longrightarrow} \mathcal{O}_{B}\left(-nK -m{S})\right)  
\longrightarrow \mathcal{O}_{{S}}\left(-nK -m{S}\right) \rightarrow 0
\, 
\label{eq:exseqbis}
\end{align}
for various integers $m,n$, where $\cdot z$ is multiplication by the defining section of ${S}$ on $B$.

As a first step, we ask that $H^1(\mathcal{O}_B(-K -{S})) \neq 0$ and $H^1(\mathcal{O}_B(-K))=0$. This guarantees that we can find a non-trivial class $\sigma \in H^1(\mathcal{O}_B(-K -{S}))$ and a section $ \tilde s_0 \in H^0(\mathcal{O}_{{S}}(-K))$ such that ${\delta}
 \tilde s_0 = \sigma$. Since $\tilde s_0$ has a non-trivial image, it cannot itself be the image of a global section on B under restriction to ${S}$. We define $\tilde f_0,\tilde g_0$ on ${S}$ in terms of $\tilde s_0$ as we discussed in section 4: $\tilde f_0 = -1/48 \, \tilde s_0^4$ and  $\tilde g_0 = 1/216 \, \tilde s_0^6$. If we further assume that $H^1(\mathcal{O}_B(-4K -{S})) = 0$ and $H^1(\mathcal{O}_B(-6K -{S})) = 0$ then exactness determines that $\tilde f_0$ and $\tilde g_0$ lift to respective global sections $f_0,g_0$ on $B$.

As in section 3, we have so far constructed an $I_1$ fiber over ${S}$, but here have avoided non-trivial monodromy (by taking $\tilde u_0 = 1/4 \, \tilde s_0^2$) with the aim of constructing $SU(n)$ fibers. It remains to give conditions for the higher order of vanishing. By construction we have 
\begin{equation}
\Delta^{(0)} := 4 f_0^3 + 27g_0^2 = {\delta _1 z}
\end{equation}
(recall $z$ is the defining section of ${S}$). Now define
\begin{align}
f^{(1)} = f_0 + f_1 z \nonumber\\
g^{(1)} = g_0 + g_1 z \, 
\end{align}
with some putative sections $f_1,g_1$. {Correspondingly, define $\Delta^{(1)} := 4 f^{(1) \, 3} + 27g^{(1) \,2}$. Vanishing at next order is the condition $\Delta^{(1)} =\delta_2 z^2$}.  This requires
\begin{equation}
\delta_1 + 12 f_1 f_0{^2} + 54 g_0 g_1 = 0 + O(z)
\, .
\end{equation} 
Restricting to ${S}$ this is equivalent to
\begin{equation}
\tilde g_1  = -\tilde f_1 {\tilde u_0} - \frac{1}{4} \tilde \delta_1 /{\tilde u_0^3}
\, ,
\end{equation}
which has a solution provided  $\tilde \delta_1 /{\tilde u_0^3}$ has no poles over ${S}$. 
Then if we require $H^1(\mathcal{O}_B(-6K-2{S}))=0$ we must have that $\tilde g_1$ lifts to a section $g_1$ of $H^0(\mathcal{O}(-6K-{S}))$ on $B$. Finding such sections we have constructed an $SU(2)$ example given by $f = f^{(1)}$ and $g = g^{(1)}$.

The general procedure follows similarly. We iteratively define
\begin{equation}
\Delta^{(i-1)} = 4 f^{(i-1) \, 3} + 27g^{(i-1) \,2} = \delta_i z^i \quad , \quad f^{(i)} = f^{(i-1)} + f_i z^i  \quad , \quad g^{(i)} = g^{(i-1)} + g_i z^i   \, .
\end{equation}
Vanishing of $\Delta^{(i)}$ at order $i+1$ requires, over {S}, that
\begin{equation}
\tilde g_i  = -\tilde f_i {\tilde u_0} - \frac{1}{4} \tilde \delta_i /{\tilde u_0^3}
\, .
\end{equation}
Moreover, $\tilde g_i$ is well-defined and lifts to some global $g_i$ on $B$ provided $\tilde \delta_i / {\tilde u_0^3}$ has no poles over ${S}$ and $H^1(\mathcal{O}_B(-6K-(i+1){S}))=0$. For an $SU(5)$ example we must successfully complete this iterative procedure for $i=1,2,3,4$ with final result the fibration $f=f^{(4)}$ and $g=g^{(4)}$.



\ifx\undefined\bysame
\newcommand{\bysame}{\leavevmode\hbox to3em{\hrulefill}\,}
\fi

\end{document}